\documentclass{ws-ijmpd}
\usepackage[super,compress]{cite}
\usepackage{amsmath,bm}
\usepackage{amssymb}
\usepackage{epsfig}
\newcommand{\npl}{np$\Lambda$}
\newcommand{\npk}{np$ K$}
\newcommand{\nplc}{np$\Lambda\Xi$}
\newcommand{\nplsc}{np$\Lambda\Sigma\Xi$}
\newcommand{\nplck}{np$\Lambda\Xi K$}
\newcommand{\MI} {$\mathbfit I$}
\usepackage{color}
\usepackage[normalem]{ulem}  
\usepackage[dvipsnames]{xcolor}

\DeclareMathAlphabet      {\mathbfit}{OML}{cmm}{b}{it}

\begin{document}

\markboth{S.S.Lenka, P.Char, S. Banik}
{Critical mass, moment of inertia and universal relations of rapidly rotating
neutron stars with exotic matter}

%
\catchline{}{}{}{}{}
%

\title{Critical mass, moment of inertia and universal relations of rapidly rotating
neutron stars with exotic matter}
\author{Smruti Smita Lenka}
\address{BITS Pilani, Hyderabad Campus, Shameerpet Mondal, Hyderabad 500078, India}
\author{Prasanta Char}
\address{Inter-University Centre for Astronomy and Astrophysics, Post Bag-4, Ganeshkhind, Pune, India}
\author{Sarmistha Banik \footnote{sarmistha.banik@hyderabad.bits-pilani.ac.in}}
\address{BITS Pilani, Hyderabad Campus, Shameerpet Mondal, Hyderabad 500078, India}
\maketitle

\begin{history}
\received{30 January 2017}
\revised{21 April 2017}
\end{history}
\begin{abstract}
\noindent  
We calculate moment of inertia of  
neutron star with different exotic constituents such as hyperons 
and (anti)kaon condensates and study its variation with 
mass and spin frequency. The sets of equation of state, generated 
within the 
framework of relativistic mean field model with density-dependent couplings are adopted for the purpose.
We follow the quasi-stationary
evolution of rotating stars along the constant rest mass sequences, that 
varies considerably with  different constituents in the equation of state.
We also explore the universal relations associated with some of the  normalised 
properties, such as critical mass and moment of inertia 
for specific EoS or
as a matter of fact constituents of the dense matter.  
Deviations in the universal relations for moment of inertia are observed 
at higher 
compactness. This study presents important results concerning the 
properties of 
neutron stars, that could be observationally verified in the near future using 
Square Kilometer Array telescope.
\ccode{97.10.Nf, 97.10.Pg, 97.10.Kc, 26.60.Kp}
\end{abstract}
\maketitle
\section{Introduction}

Out of estimated total galactic population of a few tens of thousands of 
pulsars beaming towards the earth, we 
observe $\sim2500$
radio pulsars in our galaxy so far\cite{ATNF}. Among these known sources, 
fewer than 10\% are in binary systems and only their
mass could be measured precisely using Shapiro delay or  a combination of 
post-Keplerian orbital parameters such as orbital period decay, 
advance of periastron, and constraints from spectroscopic modeling of the 
companion star \cite{Watts15}.
The best studied pulsars in the neutron
star binaries are the Hulse-Taylor pulsar, J0737-3039 etc. Masses are  
determined not only for double neutron star systems, but for several neutron 
star-white dwarf binaries as well, since they allow
measurement of orbital period decay. 
Mass of accreting neutron stars can also be constrained by
measuring the motion and spectral properties of the companions, but this 
process has large uncertainties.
 
Besides mass the other important observable of the neutron star is radius. 
There are several methods to constrain the
radius utilising  various properties of thermonuclear
bursts, accretion-powered millisecond pulsars, kHz QPOs,
relativistic broadening of iron lines, quiescent emissions, and binary orbital motions, 
but they are not free from  systematic uncertainties
\cite{Sudip10}. Independent mass and radius measurement for the same compact star 
without invoking their combination, for example red-shift, is yet to happen due 
to various uncertainties such as the estimated distance of the source, the 
unknown chemical compositions of the atmosphere, interstellar absorption, 
and the presence of magnetic field \cite {Lat_Schutz}. 

Such astronomical techniques of mass-radius measurement should be complemented 
with theoretical investigations to get a complete picture of a neutron star. 
Observationally we can not directly probe 
the deep interior of a neutron star. There are several 
microscopic models that theoretically compute the equation of state (EoS) 
i.e. pressure(P) 
and energy density($\epsilon$) at a certain density for the degenerate matter. 
Hence for a given EoS the static mass-radius relationship is generated
by solving the Tolman-Oppenheimer-Volkoff (TOV) equation. For a rotating 
star also, this mass-radius sequence can be generated upto its maximum
rotation frequency i.e. Kepler frequency.
The set of EoS that does not conform to the latest mass 
measurement\cite{demo, anton, fons} is ruled out.
However, it is equally important to constrain the EoS with the simultaneous 
observation of radius. And we lack reliable data of radius, as has already been mentioned. 
After the discovery of highly relativistic binary systems such as the double 
pulsar system PSR J0737-3039 for which masses of both pulsars are known 
accurately, it was argued that a precise measurement
of moment of inertia (\MI) of one pulsar might overcome the uncertainties 
in the determination of radius (R) since dimensionally \MI$\propto M R^2$ 
\cite{Lat_Schutz}. In relativistic binary systems, higher order post Newtonian
(PN) effects could be measured. Furthermore, the relativistic spin-orbit (SO) 
coupling may manifest in an extra advancement of periastron above the
PN contributions such that the total advance of periastron is 
$\dot\omega   =\dot\omega_{1PN} + \dot\omega_{2PN}  + \dot\omega_{SO}$ \cite{Damour88}.
The SO contribution has a small effect and could be measured when it is 
comparable to the 2PN contribution. The measurement of the SO effect leads to 
the determination of moment of inertia (\MI) of a pulsar in the double pulsar system 
\cite{Watts15, Shao15}.
However simulations assuming 5$\mu$s timing precision for the pulsar  
A of PSR J0737-3039 predict that it would 
take another 20 years to measure \MI ~at 10\% accuracy \cite {Kramer_Wex09}. 

This situation would change vastly with the advent of world\textsc{\char13}s 
largest radio telescope, 
the Square Kilometer Array (SKA), which is expected to begin 
operations early in the next decade. 
The high precision timing technique in the 
SKA would quicken the determination of the moment of inertia of the double 
pulsar PSRJ0737-3039A. 
The distinguishable effect of pulsar moment of inertia on the pulsar timing 
in SKA will be of the order 
of the 2PN level correction to the advance of periastron $\dot\omega$, 
due to highly improved precision of timing measurements over its predecessors. 
It can also measure \MI ~by estimating the spin-orbit 
misalignment angle $\delta$ in double neutron stars.  SKA
can potentially discover highly relativistic binaries, exotic system like
pulsar-black hole binary and sub-millisecond pulsar
\cite{Watts15}.
Simultaneous measurement of high mass and high spin stars would be 
particularly helpful in constraining the mass-radius relation. 
Also, the mass-radius profiles from different models can intersect among themselves. 
Therefore, we need a third 
independent parameter to constrain the EoS, i.e. spin, which is  the first 
measured quantity. So, at least three independent global parameters, 
for example, mass, radius and spin frequency, of the  same star are required in order to
constrain the EoS models, and hence to understand the dense core \cite{Sudip10}.
Moreover, the spin off from the measurement 
of \MI ~in the SKA looks promising. The first consequence is the estimation of 
radius in highly relativistic
binaries where mass of each neutron star is accurately determined. Secondly,
if the back bending effect is observed, 
it might reveal a phase transition from nuclear matter
to some exotic form of matter (hyperon or quark) in neutron star interior \cite{Haensel16}.
Furthermore, moment of inertia measurement would also shed light on the 
I-Love-Q relation. 
Masses, radii, backbending  and stability for rotating neutron stars with 
hyperon and quark cores are discussed in Ref \cite{Haensel16}. Backbending was 
also reported for rotating neutron stars with quark core in  
Ref. \cite{Gle97, Zdunik06}.
We neglect all the secular instabilities that may appear at very high rotation. 
These instabilities decay by emitting gravitational radiation,  
reducing the speed of the stars in realistic cases before attaining the Kepler 
limit. But we have not considered this in the spin down.

If the moment of inertia is measured with 10\% uncertainty, this would 
constrain the radius of a $1.4M_{\odot}$ star
with $6\%-7\%$ uncertainty \cite{Lat_Schutz}. But again, there will be a 
family of EoS models in \MI-M curve lying in that range.
Thus, it is crucial to examine the dependence of \MI ~on EoS. As of now
there are several EoS models available for neutron star core. We are yet to 
reach a consensus on the compact star EoS, due to our limited
knowledge of the nature of matter beyond normal nuclear matter density. 
Neutron star matter should encompass a wide range of densities; from the 
density of
iron nucleus at the surface of the star to several times normal nuclear matter 
density in the core. At this high density interior of neutron stars, the 
chemical potentials of nucleons and leptons increase rapidly. Consequently 
several novel phases with large strangeness fraction such as, hyperon matter, 
Bose-Einstein condensates of strange mesons and quark matter may exist. The 
presence of exotic matter typically softens the EoS,  
resulting in a smaller maximum mass neutron star than that of the nuclear EoS 
\cite{Gle}. So, does the deep interior of a neutron star really contain the 
strangeness degrees of freedom? 
Comparing the values of mass-radius obtained using the astronomical techniques 
along with the theoretical predictions may solve the long-standing puzzle. 
Currently the accurately measured high-mass neutron stars are PSRs J1614-2230 
and J0348+0432 with masses $1.928\pm 0.017$ and $2.01 \pm 0.04\: M_{solar}$ 
respectively {\cite{demo, anton, fons}. They put a strong 
constraint on the EoS of neutron star matter and rule out most of the soft 
EoS only from the mass measurement.  

In recent years, there has also been an alternative approach to this puzzle, 
namely the study of universal relations among several quantities that 
characterizes a compact object. These relations are supposed to be independent 
of the EoS of the star. Therefore if one of them is measured, the other
 can be estimated from an analytical expression without taking into 
consideration the internal structure of the star. It has been discovered that 
such relations exist among the moment of inertia (\MI), the Love number 
associated with the tidal deformation and spin induced quadrupole moment ($Q$)
 \cite{yagi,yagi13}. 
But, many of these universality relations are approximate as they exist only in some particular regime such as 
slow rotation approximation and not very strong magnetic field. It has been 
shown that at very high rotational frequencies the universality between
\MI ~and $Q$ breaks down as both the deviations due to EoS and spin frequency 
are comparable at high spin frequency \cite{doneva14}. It has also been shown 
that a comparatively slowly rotating star ($P \gtrsim 10s$) and strong magnetic 
fields ($B \gtrsim 10^{12}G$),
the universality between \MI ~and $Q$ breaks down \cite{haskell14}. 
The slowly rotating stars are found to abide by universal relations as 
reported  in Refs.\cite{yagi,yagi13}. 
Many authors tried to highlight a relation between normalised moment of 
inertia (\MI$/MR^2$) and stellar compactness($M/R$)
 \cite{Lat_Schutz, Raven94, Lat01, Breu, doneva14, sayan14}. Also, having a 
universal relation 
will allow us to determine the radius with a very high accuracy 
once we have simultaneous measurements of \MI  ~and mass of the same star. 
 
Recently, Breu and Rezzolla \cite{Breu} have studied such universal relations
in great detail. They have also used another 
normalisation for moment of inertia as (\MI$/M^3$) and showed that universality
relation holds more tightly than the previous case. In their analysis, 
they have taken a large set of EoS with different stiffness. 
However, the matter they considered is nucleonic. They satisfy the two solar mass limit, 
but many of them use non-relativistic interactions, and some of the 
relativistic EoS also use the parameter sets that are not favorable in view of 
the symmetry energy nuclear experimental data. 
It is not known 
if the inclusion of exotic components or a phase transition inside the neutron 
star core would effect the universality relations. This is one of 
the main goals of the present work.
We are also motivated to study the effect of higher rotational frequencies on 
the universality relation between normalized moment of inertia and stellar 
compactness. 

Here in this paper, 
we consider different compositions in our 
EoS which is generated under the framework of density dependent  relativistic
mean field model \cite{dd2_14,ddrh} and study the variation of mass-radius 
profile for different EoS and how they evolve with uniform  rotation of the 
neutron stars. We also study the maximum mass sustainable for rotating 
configuration, expressed in terms of dimensionless and normalised angular 
momenta. Next, we follow the variation of \MI ~with 
gravitational mass and 
observe how rotation affects \MI ~for different constituents of dense 
matter relevant to the neutron star core. 
Finally, we investigate if there exists a 
universal relations 
between \MI ~and compactness (M/R). We use two types of 
normalised moment of inertia relations prescribed by Ref
\cite{Lat_Schutz, Breu} for EoS with nucleon, hyperon and antikaon condensates 
degrees of freedom 
for rapidly rotating compact stars. 

The paper is organised as follows. In Section II, we briefly discuss the
rotating neutron stars, the rns code that we use to generate our results
and EoS. Section 
III is devoted to results and discussion. Finally we summarise in 
Section IV. 

\section{Rotating neutron stars and EoS}

Pulsars are rotating neutron stars with a strong magnetic field. Their magnetic 
axes are not aligned to their rotation axes and the continuous radio emission 
from the magnetic poles sweeps our line of sight periodically. 
Observations convey that the time period of pulsars vary from 
approximately 1 ms to  10 seconds \cite{ATNF}.  Most of the pulsars discovered so far, 
take nearly 1 sec to complete one revolution. The slowest one, observed in 2011,
takes an exceptionally large amount of time (1062 s) compared to the rest.
The pulsar PSR J1748-2446ad with a frequency of 716 Hz is the fastest. 

A rotating neutron star is composed of uniform matter of degenerate baryons in the 
core where neutrons and protons could be superfluid and superconducting 
respectively, surrounded by a crust made of nuclei and superfluid neutrons and
is modeled by a stationary, axisymmetric, perfect fluid space-time.
The presence of solid 
crust contributes to  negligible departure (of the order of $10^{-5}$) from 
perfect fluid equilibrium. 
Within the few years of formation, 
the interior temperature of a neutron star becomes $\sim 10^9 K$, that is 
negligible compared to the Fermi energy of the constituent dense matter.
Hence, the effect of finite temperature is ignored. 
Also with time, the outer crust becomes superfluid and the nucleons form an array 
of vortices due to rotation.  The characteristic length over which the gravitational 
field of the rotating star varies is much larger than distance between the 
vortices. So the relativistic star is modeled as a uniformly rotating, 
zero-temperature perfect fluid\cite{Friedman}. 

As there is no analytical self-consistent solution for the space-time,  
several numerical codes are developed to study the rotating relativistic stars.
We use rns code \cite{Str}, which constructs models of rapidly rotating, relativistic,
compact stars using  Komatsu, Eriguchi and Hachisu (KEH) scheme\cite{KEH}. Here the field 
equations and the equation of hydrostatic equilibrium are solved iteratively 
by fixing the central energy density and any of the other variables such as
mass, rest mass, angular velocity, angular momentum or the ratio of polar 
radius to the equatorial radius until convergence. We run the code for our 
tabulated, zero-temperature EoS, consisting of energy density, pressure, 
enthalpy and number density. The code 
also incorporates modifications by Cook, Shapiro and Teukolsky \cite{CST}, where
a new radial variable is introduced that maps semi-infinite region to the 
closed region, unlike KEH scheme. In KEH scheme, the region of integration 
was truncated at a finite distance from the star. 

A rotating star can support larger mass compared to a static one.
In a rapidly rotating neutron star model, we can follow the quasi stationary 
evolution of a single star along the constant rest mass sequences.
We have two equilibrium sequences termed as "normal" and "supramassive".
Normal sequences start rotating with Keplerian frequencies and end up 
on the static limit, which is stable to quasi-radial perturbation. On the
other hand, the supra-massive sequence does not have a 
corresponding static solution. They have rest masses that are higher than 
their static (TOV) counterparts. We study these high mass stars in this work.

Recently, we adopted a density dependent relativistic mean field model where 
hyperons and antikaon condensates were considered apart from the nucleons 
and leptons \cite{dd2_14}. In this model baryon-baryon interactions are 
mediated by 
$\sigma$, $\omega$ and  $\rho$ mesons. The salient feature of this model is 
the appearance of rearrangement
term due to density dependent meson-baryon couplings which takes care of many
body effects and the thermodynamic consistency \cite{PRC87} in dense 
neutron star matter. Nucleon-meson couplings of the model
are determined from binding energies, spin-orbit splittings, charge and
diffraction radii, surface thickness and neutron skin of finite nuclei
following the DD2 model of Typel {\it et. al} \cite{typel05,typel09}.
Consequently, our model leads to symmetric nuclear matter properties at the
saturation density for example binding energy per nucleon 16.02 MeV,
incompressibility 243 MeV, symmetry energy 31.67 MeV and its slope parameter
corresponding to the density dependence of symmetry energy 55.03 MeV. These
values, particularly symmetry energy and its slope parameter are in excellent
agreement with experimental findings \cite{Lim}.
The equation of state (EoS) in the sub-saturation density regime is well 
constrained. 
It should be noted that other density dependent parametrizations were
extensively used in the calculation of neutron stars for example DDME2 
parameter set \cite{Ring}. It has been shown that
nuclear matter properties at the saturation density and maximum neutron star
mass calculated with DD2 and DDME2 sets are very similar \cite{PRC94}. 

The density-dependent meson-hyperon couplings are 
obtained from the density dependent meson-nucleon couplings using hypernuclei 
data \cite{Sch} and scaling law \cite{Kei}. Analysis of 
hypernuclei events predict an attractive potential for $\Lambda$ and $\Xi$ and 
a repulsive potential for $\Sigma$ in symmetric nuclear matter. We have considered 
$\Lambda$, $\Sigma$ and $\Xi$ hyperons in this calculation  
and the scalar meson couplings to these hyperons are fitted 
to the experimental value of potential depth of 
respective hyperons ($U_{\Lambda}^N(n_0)=-30$ MeV, $U_{\Xi}^N(n_0)=-18$MeV, 
$U_{\Sigma}^N(n_0)=+30$ MeV) in saturated nuclear matter \cite{dd2_14}.
The repulsive interaction among the hyperons are also considered within the 
model, it is mediated by the exchange of $\phi(1020)$ mesons. The 
couplings of antikaon-nucleon interactions are obtained similarly, 
however, they are not density-dependent. 
The particular choice of hyperon-nucleon potential does not affect 
the maximum mass of neutron stars \cite{weiss1}, but it varies with optical
potential of (anti)kaons in nuclear matter \cite{dd2_14}. Recently there have 
been studies of hypernuclear matter in neutron stars which meet the 
$2M_{solar}$ mass limit \cite{PLB14, ApJ17}.  The ref \cite{ApJ17} gauges the 
parameters of the density-dependent functional with hypernuclei. There are 
several hyperon EoS 
which are reviewed in Ref \cite{Fortin15}. Out of them presence of $\Sigma^{-}$ for a repulsive potential is 
confirmed along with $\Lambda$ and $\Xi$ hyperons only in case of a few EoS 
models,
i.e. BM165 \cite{Bednarek12},  DS08 \cite{Dex08}, UU1 and UU2\cite{Uechi}.
However, only BM165 and DS08 are 
consistent with $2M_{solar}$ observation\cite{demo,anton,fons}.

Our EoS including hyperon and antikaon condensed matter is 
consistent with the 
observational limit of $2M_{solar}$\cite{demo,anton,fons}, unlike most of the 
existing exotic EoS. We restrict our comparison to hadronic EoS only.
Furthermore, it is interesting to note that the pure neutron matter EoS 
calculated with DD2 set agrees well
with the Chiral effective field theory result\cite{hempel14,RMP17}.

\section{Results}

We report our results for rotating neutron star calculated using rns code 
\cite{Str} for our EoS.
The fastest rotating pulsar observed until now has a frequency $716$ Hz or angular velocity $\Omega\approx4500 s^{-1}$. 
Our choice of angular velocity $(\Omega)$ values
($5300, 5800~ \& ~6300 s^{-1}$) 
are slightly more but of the same order of magnitude. The motivation
of these choices are to study the effect of rapid rotation on the universal 
relations in case of millisecond or sub-millisecond pulsars. 
The different sets of EoS, Pressure ($P$) versus energy density 
($\epsilon$), are plotted in Fig 1 and denoted by their 
respective compositions.  In the Fig 1a, we plot the 
nucleons-only (np) EoS in red; 
neutron, proton, $\Lambda$ (\npl) EoS in blue; neutron, proton, $\Lambda$, $\Sigma^-$, $\Xi^0$ (\nplsc ) EoS in cyan; 
neutron, proton, 
$\Lambda$, $\Xi^0$, $\Xi^-$ (\nplc) EoS in green and neutron, proton, $\Lambda$, $\Xi^0$, 
$\Xi^-$, antikaon condensates(\nplck) EoS in maroon.
These EoS are generated within the framework of 
DD2 model \cite{dd2_14}. 
Presence of hyperon makes the EoS softer compared to nucleons-only case.
For \npl ~EoS $\Lambda$ appears at $2.20n_0$, for \nplsc ~EoS $\Sigma^-$ and 
$\Xi^0$ appear 
at $2.48n_0$ and $6.26n_0$ respectively. Presence of $\Sigma^-$ does not allow 
$\Xi^-$ in  the system.
Finally for \nplc, $\Xi^-$ and $\Xi^0$ appear at $2.44n_0$ and $7.93n_0$ 
respectively\cite{dd2_14}.
The study of Kaonic atoms suggests an attractive
optical potential for the (anti)kaons in nuclear matter. 
However, there is no definite consensus how deep the potential is. 
We report our results for a set of values of 
$U_{\bar K}$ from -60 to -140 MeV in Fig 1b. These EoS are for neutron, proton 
and antikaon condensates (denoted by \npk).
The coupling constants for kaons with $\sigma$-meson, $g_{\sigma K}$ 
at the saturation density for the range of $U_{\bar K}$ 
for DD2 model are listed in Table 2 of Ref \cite{dd2_14}. 
The deeper the antikaon potential in nuclear medium, softer is the EoS.
In Fig 1a we also consider the 
presence of $\Lambda$ and $\Xi$ hyperons as well as the (anti)kaon condensates.
Since early presence
of hyperons delay the appearance of $K^-$ condensates to higher density, effect
of $K^-$ condensates is only considerable for $|{U_{\bar K}}|\geq\: 140$ MeV.
So we consider neutron, proton, {$\Lambda$}, {$\Xi$}, antikaon condensates (\nplck) 
for $U_{\bar K} =-140$ MeV only in Fig 1a.

\begin{table}[ph]
\tbl {Maximum mass (in $M_{solar}$) 
 and the corresponding radius(in km)  of compact stars with 
nucleons, hyperons and (anti)kaons in DD2 model. 
$U_{\bar K}$ =-140 MeV for $np\Lambda\Xi K$ matter. 
The values in the parentheses are for the Kepler sequences.}
{\begin{tabular}{@{}ccc@{}} \toprule
&$M_{static[Kepler]}$ ($M_{solar}$)&$R_{static[Kepler]}$ (Km)\\
&($M_{solar}$)& (Km)\\ \colrule
np&2.42[2.91]&11.91[15.68] \\
np$\Lambda $&2.10[2.52]&11.55[15.94]\\
np$\Lambda \Sigma \Xi$&2.07[2.48]&11.56[15.82]\\
np$\Lambda \Xi$&2.03[2.43]&11.49[15.97]\\
np$\Lambda \Xi  K $&2.01[2.41]&11.49[16.02]\\ \botrule
\end{tabular} \label{tab1}}
\end{table}

In Fig 2a and 2b, we plot the gravitational mass as a function of equatorial 
radius for different compositions (np, \npl, \nplsc, \nplc, \nplck, denoted by 
red,  blue, cyan, green and maroon respectively in the online version). The 
gravitational mass of neutron star for 
static  sequence and the  sequence rotating at their Kepler frequencies are 
represented by solid lines at the left and right extremes. The solid circles 
on these sequences indicate the maximum masses for the sequences. 
We have noticed in Fig 1a, the overall 
EoS is softer as we increase the degrees of freedom in the form of exotic 
particles. This leads to fall of maximum mass values from np 
to \nplck  ~in both 
the static and mass shedding sequences. However, the corresponding radius does 
not change much with the composition. 

\begin{table}
\tbl{Maximum mass (in $M_{solar}$) and the corresponding radius(in km) 
of compact stars with 
nucleons and (anti)kaons for different values of
optical potential depth in the DD2 model. 
The values in the parentheses are for the Kepler sequences.}
{\begin{tabular}{@{}ccc@{}} \toprule
$U_{\bar K}$(MeV)&$M_{static[Kepler]}$ &$R_{static[Kepler]}$ \\
&($M_{solar}$)& (Km)\\ \colrule
-60&2.37[2.88]&12.19[16.10] \\
-80&2.34[2.85]&12.20[16.38]\\
-100&2.29[2.80]&12.18[16.45]\\
-120& 2.24[2.72]&12.05[16.47]\\
-140&2.16[2.62]&12.03[16.27]\\ \botrule
\end{tabular} \label{tab2}}
\end{table}

The maximum mass and corresponding radius of the
star with different composition for static and Kepler sequences are listed in 
Tables \ref{tab1} and \ref{tab2}.
We notice that a star rotating at its maximum possible frequency i.e. Kepler 
frequency, can support more mass compared to the static one. As the 
centrifugal force increases with the rotational velocity, the stars tend to 
have larger radii. For each set of EoS the maximum mass increases almost by 
20\% from static to Kepler. 
In between static and Kepler sequences, three fixed angular velocity sequences 
are plotted in Fig 2a for $\Omega = 6300, 5800,$ and $5300\: s^{-1}$ 
denoted by dotted, dashed and dash-dotted line respectively.
The lines with different constituents but same frequency converge for 
relatively larger stars, whereas for smaller stars different frequency lines
merge for same set of constituents. Or in other words, mass sequence of the
small(large) rotating stars are dependent(independent) of EoS. 

In between the static and Kepler limits, fixed rest mass curves are drawn in  
Fig 2b.  
The horizontal curves denoted by I, II, III and IV are 
for rest mass, $M_R=$ 1.92, 2.49, 2.75 and 3.23 $M_{solar}$ respectively. 
All the lines in I,  np (red) in II and III are normal 
sequence; np (red) in IV, \npl (blue), \nplsc (cyan), \nplc (green) 
and \nplck (maroon) lines in II and III are supramassive.
The red dashed line marked as IV is for $M_R=3.23M_{solar}$. This value is 
well above the maximum mass for the exotic EoS, here 
we have a single sequence corresponding to np EoS only. 
Stars on the supramassive sequence may evolve keeping their 
rest mass constant but changing the spin rate. The stars on the supramassive 
sequences typically lose energy and 
angular momentum very slowly by emitting radiation. As they reach the point of 
quasi-radial instability, they rapidly spin up before collapsing into a black 
hole \cite{CST94}.

We plot the critical masses i.e., the maximum masses(M) of the constant angular 
momentum(J) stellar models, such that $(\partial M/\partial \rho_c)_{J}= 0$, 
where $\rho_c$ is the central density. 
Beyond this point the stellar models become unstable. The 
critical masses vary considerably when plotted with the angular momentum as 
seen in 
Fig 3a. Each sequence starts at $M_{crit}=M_{static}$ where $M_{static}$
is the maximum mass of a non-rotating sequence and ends at  
maximum angular momentum $J=J_{Kepler}$ with $M_{crit}=M_{Kepler}$. They vary 
for different EoS. Larger mass of stellar configurations, for example np, 
supports higher Kepler frequency or angular momentum as can be seen from 
Fig 3a. However, the percentage change of stable mass configuration as the 
static star spins up to its Kepler limit is similar($\sim 16-20\%$) for all the EoS. So we draw
the same plots in Fig 3b, but in terms of dimensionless quantiles 
$M_{crit}/M_{static}$
and normalised angular momentum where, $j=J/M_{crit}^2$ and 
$j_{Kepler}=J_{Kepler}/M_{Kepler}^2$.  
We observe that the relation between normalized critical masses 
and normalized angular momentum does not vary much with the 
given equations of state. We find a universality
(a  $20\%$ rise of $M_{crit}$ value over $M_{static}$ for all the EoS) 
in the normalised mass-angular 
momentum profile. We got a best-fit line 
$M_{crit}/M_{static}= 1 + a_1 x +a_2 x^2 + a_3 x^3$, where  $x=j/j_{Kepler}$ 
and the coefficients are $a_1= 0.3363$,
$a_2 = 0.3829$, $a_3 = 0.138$. 
This treatment is particularly useful to compute the maximum mass 
configuration allowed by uniform rotation in terms of its corresponding 
static sequence for all the EoS.
Similar features have been reported in 
Ref \cite{Breu} for nucleonic EoS, though their best fit parameters are 
slight different from ours. The dashed line with their parameters does not fit 
our data well at lower $j/j_{kep}$. 

\begin{table}
\tbl{Maximum moment of inertia and corresponding 
masses, radii and angular velocity of compact stars with 
nucleons, hyperons and (anti)kaons 
in the DD2 model.  \MI ~is in  $10^{45}  g\: cm^2$, 
maximum mass is in $M_{solar}$, radius in km 
and rotational velocity $\Omega$ in $10^4  s^{-1}$,
$U_{\bar K}$ =-140 MeV for $np\Lambda\Xi K$ matter.}
{\begin{tabular}{@{}ccccc@{}} \toprule
&\MI &M &R &$\Omega$ \\ 
&($10^{45}g\: cm^2$)&($M_{solar}$)&(Km)&($10^4\:s^{-1}$)\\ \colrule
np&5.70&2.84&16.8&0.874\\
np$\Lambda $&4.53&2.43&17.2&0.784\\
np$\Lambda \Sigma \Xi$&4.45&2.38&17.4&0.766\\
np$\Lambda \Xi$&4.30&2.32&17.5&0.752\\
np$\Lambda \Xi  K$&4.30&2.32&17.5&0.751\\ \botrule
\end{tabular} \label{tab3}}
\end{table}

In Fig 4a we plot the variation of \MI ~versus gravitational mass for EoS with 
different constituents at the Kepler frequency. 
For a neutron star of known mass and \MI, we can 
deduce the radius using  \MI ~versus M and M ~versus R graphs.  
This eventually  would
help us constrain the EoS and confirm if the NS 
contains exotic core or not. In Fig 4a \MI ~values are plotted for np, \npl,
\nplsc, \nplc  ~and  \nplck ~($U_{\bar K}= -140$ MeV)  
as well as for \npk; $U_{\bar K}$ varies from -60 to -140 MeV.
It is noted that \MI ~is maximum for stiffer EoS, i.e. for 
np, it can support a heavier and more compact star.  
The maximum of \MI ~curve however does not 
correspond to the maximum neutron star mass and the corresponding radius.
The values of maximum  \MI ~and 
the corresponding masses, radii and angular velocity are listed in 
Tables \ref{tab3} and 
\ref{tab4}. Currently, 
for PSRJ1614-2230, the lower limit of \MI ~is estimated to be 
$~10^{45}g\: cm^2$ 
from $\gamma$-ray flux measurement \cite{Abdo09}. This value is not very 
constraining due to uncertainty in the distance of the pulsar. The accuracy
is expected to improve highly with the upcoming SKA results. 

In Fig 4b, \MI ~is plotted for different angular frequencies 
($\Omega$=5300,5800 and 6300 $s^{-1}$). We have considered np, \npl, ~\nplsc, 
~\nplc 
~and \nplck ~EoS. \MI ~is less for a star rotating at $\Omega=5300s^{-1}$ 
compared to that of $6300s^{-1}$.

\begin{table}
\tbl{Maximum moment of inertia and corresponding  masses, radii and angular velocity of compact stars with
nucleons and (anti)kaons for different values of
optical potential depth in the DD2 model. \MI ~is in  $10^{45} g\: cm^2$,
maximum mass is in $M_{solar}$, radius in km
and rotational velocity $\Omega$ at $10^4  s^{-1}$. }
{\begin{tabular}{@{}ccccc@{}} \toprule
&\MI &M &R &$\Omega$ \\ 
&($10^{45}g\: cm^2$)&($M_{solar}$)&(Km)&($10^4\:s^{-1}$)\\ \colrule
-60&5.70&2.84&16.8&0.874\\
-80&5.68&2.81&16.94&0.860\\
-100&5.59&2.74&17.22&0.832\\
-120&5.37&2.67&17.33&0.814\\
-140&5.01&2.53&17.5&0.780\\ \botrule
\end{tabular} \label{tab4}}
\end{table}

The moment of inertia versus the angular velocity for all np, 
\npl, \nplsc,
\nplc ~and \nplck ~cases are plotted in Fig 5a. Here also the same colour 
scheme 
is used for different compositions as in Fig 1. 
We find that \MI ~is lower for softer EoS (\nplck) 
compared to the stiffer ones (np, \npl ~etc) at a particular angular frequency. 
The four solid lines represent \MI ~at Kepler frequency, each of those 
increase to a peak but falls for a relatively slower pulsar.  
For stiffest EoS, np, \MI ~peaks at 
$\Omega=8740s^{-1}$, while  the maximum occurs at $\Omega=7840s^{-1}$, $7660s^{-1}$,
$7520s^{-1}$, $7510s^{-1}$  for  \npl, \nplsc, \nplc ~and \nplck ~($U_{\bar K}= -140 MeV$) respectively (See Table \ref{tab3}).
We notice that the maximum angular velocity of 
rotating neutron stars is quite sensitive to the EoS. An EoS
predicting smaller Keplerian frequencies than the observed frequencies can be 
ruled out and thus be useful in constraining EoS. We also plotted fixed rest 
mass $M_R$ sequences in the same figure, which show different behaviour for 
different $M_R$ and EoS. To understand the nature of moment of inertia curves 
in Fig 5a, we 
studied the variation of mass and radius terms with respect to angular 
frequency 
We mention these results only qualitatively to analyse Fig 5a (not shown in 
separate graphs). 
For $M_R=1.92 M_{solar}$, the sequences are found at lower $\Omega$ region. 
Stars that are less massive can not sustain rapid rotation. We have noted 
both the mass and radius decrease 
monotonically at this angular frequency range, which explains the 
nature of \MI ~curves. 
Also, The variation of \MI ~with different compositions is negligible in this 
case. 
This is quite expected as we have found
in horizontal curve I of Fig 2b that all EoS merge together for 
$M_R=1.92M_{solar}$.
The central energy density of a $1.92M_{solar}$ static star differs from 
7 to $7.96\times10^{14}g/cm^3$ for np to \nplck, 
whereas hyperons start populating at $5.91\times10^{14}g/cm^3$. 
The corresponding central density for a $M_R=1.92M_{solar}$ 
star rotating with Kepler frequency is only $5.9\times10^{14}g/cm^3$, so the 
effect of 
exotic particles is not there. The supra-massive star sequence 
$M_R= 2.49M_{solar}$ also falls monotonically as $\Omega$ decreases in case of 
all EoS(long dashed lines).
However the monotonic fall is followed by a sudden spin-up with a slight 
loss of \MI ~for this supra-massive sequence when the star contains exotic 
components.
The nature of bending is different for different EoS and
can be justified by following the variation of mass and radius with $\Omega$. 
Both of them decrease initially with decreasing $\Omega$ to rebound 
at a particular angular frequency suddenly.
Along a supra-massive sequence of a pulsar as $\Omega$ decreases,
a spin up followed by a second spin down after the initial one  
in the \MI ~versus $\Omega$ ~plane was reported in Ref \cite{Weber99,Zdunik04,Banik04}.
This phenomenon is known as the back bending (S-shaped curve in the plot) 
and was attributed
to the phase transition from nuclear matter to some exotic 
(hyperon, antikaon condensed or quark) matter.
Interestingly, for the supra-massive sequence ($M_R=2.75 M_{solar}$, the 
dotted curves), value of \MI ~falls off with higher angular velocity for the exotic EoS.  
However, $M_R=2.49$ as well $2.75 M_{solar}$, corresponding to red long dashed 
and dotted lines, are normal sequences for np EoS 
and follow the same characteristic fall with decreasing $\Omega$ ~like the 
normal sequences ($M_R=1.92 M_{solar}$). 
Stars at normal sequences do not exhibit 
spin up even if they lose energy and angular momentum. This was also observed  
by Cook 
{\it et. al} in Ref \cite{CST94}.  
The red dot dashed line is supramassive for np ($M_R=3.23 M_{solar}$) and follows the same 
nature as 
the other exotic supramassive sequences with $M_R=2.75 M_{solar}$. 
For supramassive sequences, we also 
notice that the size of the star increases at low rotation speed. Though 
it can support a higher mass at higher rotation speed like the normal 
sequence, at low speed a higher mass star can be supported again. 
We show the variation of $\Omega$ with the 
angular momentum $J$ in Fig. 5b for the same sets of parameters as in Fig 5a.
This helps us to check the stability of the rotating configurations 
\cite{Zdunik04}. The condition for instability
along a constant rest mass sequence is $\frac {dJ}{d\epsilon} \geq 0$ 
\cite{Gle}.
Supramassive sequences terminate at different values of $\Omega$ 
below which the instability appears. It is found that angular momentum 
initially decreases with central energy density ($\epsilon$). After the 
stable region, angular momentum (J) starts increasing with central energy 
density ($\epsilon$). We notice the increase of J as the 
star spins up subsequently. This marks the instability in the 
configurations with respect to axisymmetric perturbations. 

The dependence of \MI$/MR^2$ with compactness (M/R) is drawn in Fig 6a. It was 
shown 
by Lattimer and Schutz that other than very soft EoS, which gives a maximum mass
of the order of $1.6M_{solar}$, a relatively unique relation existed
between \MI$/MR^2$ and M/R \cite{Lat_Schutz}. We however note that though
at lower compactness, \MI$/MR^2$ is independent of EoS, it 
clearly varies differently for different EoS at higher 
compactness; 
it is quite less for the soft EoS compared to nucleons-only EoS. 
\MI$/MR^2$ term on the other hand depends on  
angular velocity $\Omega$ at lower compactness. 
It is less for a star spinning at higher rotational speed. 
The dependence on $\Omega$ is not that prominent at higher compactness.
At this point we recall that 
all our EoS yields a maximum mass not less than $2M_{solar}$ (See 
Tables \ref{tab1}, \ref{tab2}). Still
we find deviation from universality at higher compactness for different constituents. We notice at 
low compactness, which is near Kepler frequency,
the central energy density is such that hyperons or antikaons do not populate. 
So all the EoS are identical. Naturally here \MI$/MR^2$ is independent of EoS. 
However at 
higher compactness when the central energy density is high enough to populate 
exotic particles, \MI$/MR^2$ varies differently for different EoS. The deviation
for a particular compactness(say 0.25-0.27) varies between 5-7\% from np to
\nplck ~EoS. If we consider the highest Keplerian frequency, the difference
between np and \nplck ~EoS can be as high as 14\%. 

Similar trend is noticed for rapidly rotating stars from Fig 6b also:
\MI$/MR^2$ for \npk ~at different $U_{\bar K}$, is independent of EoS only at 
low compactness. 
The difference of \MI$/MR^2$ for softest and stiffest EoS at 
Keplerian frequency is $\sim 7$\%, whereas for a compactness 0.27, this 
varies around 2.5\% 
as we change the optical potential of antikaons in 
nuclear matter from -60MeV to -140MeV. 
On an average 
these deviation from universality is of the same order as reported by Breu and 
Rezzolla for nucleonic EoS in Ref \cite{Breu}. 

We exhibit variation of moment of inertia in terms of another dimensionless 
quantity, \MI$/M^3$ with compactness for np, \npl, \nplsc, \nplc, 
\nplck ~in Fig 7a \cite{Breu}. The three distinct bunches of lines differ in 
their angular 
velocity values $\Omega$, increasing from top to bottom. 
It is clearly seen 
that  \MI$/M^3$ is also not affected by the different EoS and respects 
universality relation at lower compactness. 
However at higher compactness, a slight deviation from 
universality is evident. This part is zoomed in the inset. At a particular 
M/R value, \MI$/M^3$ drops considerably from np to \nplck ~EoS.

In the next graph (Fig 7b) we look into the variation of  \MI$/M^3$ with 
compactness for \npk ~case. Here we consider different optical potential
of antikaons  $U_{\bar K}=  -60$ to $-140$ MeV. The effect is similar as in 
Fig 6b. The variation of \MI$/M^3$ is dependent on 
$\Omega$ values at lower compactness, it decreases for a faster rotating star. 
The variation with different composition i.e., EoS 
at higher compactness is clearly visible in the zoomed inset.

Finally in Fig 8, we show the variation of the normalised moment of inertia for the 
rest mass sequences I, II, III and IV of Fig 2b. 
Fixed rest mass sequence has also been used by Martinon et al. \cite{martinon}
to study quasistationary evolution
 of I-Love-Q universal relations for a particular EoS.
In Fig 8a, we note  \MI$/MR^2$
changes appreciably with EoS for supra-massive sequences, however is 
practically independent of EoS for 
normal sequence ($1.92M_{solar}$ rest mass). The same pattern is noted for
\MI$/M^3$ in Fig 8b. The anomalous behavior between the normal and supramassive 
sequences in Fig 8a and 8b, may be explained as the difference between the 
evolutionary track of an isolated star of sequences. While the normal sequence 
stars never spin up, the ones in the supramassive sequence spin up differently
depending on their EoS. The universality is a signature of a balance between 
gravity and the response of matter to gravity. These supramassive sequences 
seem to disrupt the balance. 

\section{Discussion and Conclusion}

Our main objective is to study the moment of inertia and different universal 
relations associated with it for a neutron star that 
has several exotic particles such as hyperons($\Lambda$, $\Xi$, $\Sigma$), 
(anti)kaon condensates in its high density core.
We have constructed a set of EoS incorporating these exotic components  
individually as well as considering all of them together in the framework of 
a density dependent hadronic field theory using the DD2 parameter set. 
All our EoS satisfy the $2M_{solar}$ observational constraint. The 
EoS of hyperon matter calculated using DDME2 set also resulted in 2 M$_{solar}$
\cite{PLB14,PRC87}.}   
We employ them to study the structure of rapidly rotating neutron stars. 
Also as the stars spin up, the maximum masses they can support go up and the 
stars become larger in size  
for all the EoS considered.
In this connection, we consider the equilibrium mass of the star at maximum 
angular momentum, beyond which the configuration becomes unstable and call 
this mass as critical mass following Ref \cite{Breu}. We find that though 
different equations of state give rise to different maximum masses and 
end up at different angular momenta, the
maximum mass supported by rotation is almost 16-20\% more than that 
supported by 
their corresponding static sequences for all the EoS. Hence we
express this critical mass sequences of rapidly rotating stars
in terms of the static mass solution and normalised angular momentum 
and show that a universal relation holds irrespective of the different 
constituents  or EoS, the variation at Kepler limit being 20\% larger than the 
static limit. This number is in full agreement with the results for nucleonic 
EoS  by Breu and Rezzolla \cite{Breu}.

Next we focus on moment of inertia for different compositions and explore 
its variation with mass. 
The larger the value of \MI, the larger is the 
maximum mass supported by the stellar models, and more compact is the star.
Also, stiffer EoS  can withstand higher angular velocity. We came to  these
conclusions for the dense matter containing exotic degrees of freedom. 

We followed the fixed rest mass  sequences of stellar models, both normal and 
supramassive and studied the variation of \MI. Along normal
sequences, \MI ~falls off monotonically as the stars spin down. However, for 
the supramassive cases, we notice sudden spin up of the stars following a spin down.
\MI ~drops drastically during the spin down, but the fall of \MI ~is
not so prominent during the 
subsequent spin up. This effect was maximal in the  \npl ~case.
Next, we study the universal relations for normalised \MI. They are very 
useful tools in astrophysics.  With the upcoming SKA telescope, 
measurement of \MI ~may be accomplished soon,  which will help constraining 
the EoS. However, it is incredibly difficult because of a few things. 
First of all, there are very few relativistic binaries where the effect of 
\MI ~on the periastron advance at 2PN can be
observed. Then, in addition to periastron advance, at least two other 
post-Keplerian parameters are
needed to be measured precisely. Even if the effect of spin orbit coupling 
is strong enough, we need to
accumulate data of periastron advance for a minimum 20\% time of the precision 
period of the pulsar
to get a minimum accuracy of 10\% in the moment of inertia measurement. The 
precision period of
one of the most suitable candidates for this measurement, PSR J0737-3039, 
is 75 years. 
Hence, both mass and \MI ~measurements are needed to be 
combined with the universal relations 
to get more accurate estimation of the radius as emphasized also by Breu and 
Rezzolla \cite{Breu}.
That is why it is vitally important to know if universality relations hold for the equations of state with
considerable softening at higher densities in the presence of hyperons and antikaon condensates etc.

Therefore, we have studied the variation of normalised
\MI ~with respect to compactness (M/R) for all the constituents np, \npl, \nplsc, \nplc,
\nplck ~($U_{\bar K}=-140$ MeV) and \npk ($U_{\bar K}=-60$ to $-140$ MeV).
\MI ~is normalised with respect to $MR^2$ as well as $M^3$. 
We report a 10\% deviation on an average from universality at higher 
compactness for all the equations of state.
However at lower compactness, the normalised \MI ~is independent of our choice 
of composition, only vary with angular velocity. 
We have also investigated the normalised \MI ~vs compactness relations for fixed rest mass sequences. 
The variation is quite large for supramassive sequences, while practically
insensitive for the normal sequences.
Thus, we conclude that except from the supramassive sequence stars, the universality relations holds true for a normal star with exotic components, 
as the deviation from universality is of the same order for stars with only 
nucleonic EoS previously reported.
 
\section{Acknowledgement}
S. S. Lenka would like to acknowledge the support of DST, India through INSPIRE fellowship. The authors are also thankful to Debades Bandyopadhyay for 
carefully reading the manuscript.

\newpage 
\begin{figure}
\begin{minipage}{.5\textwidth}
  \includegraphics[width=.8\textwidth, height=0.5\textheight]{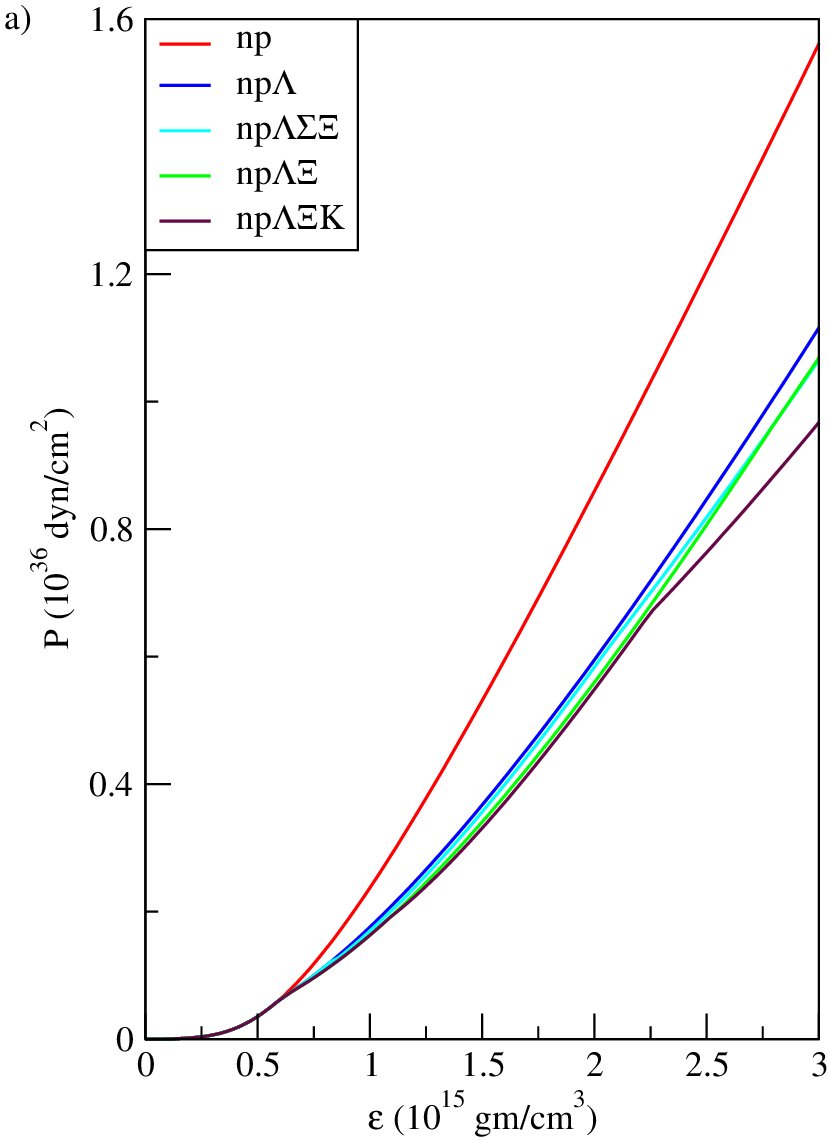}
\end{minipage}%
\begin{minipage}{.5\textwidth}
  \includegraphics[width=.8\textwidth,height=0.51\textheight]{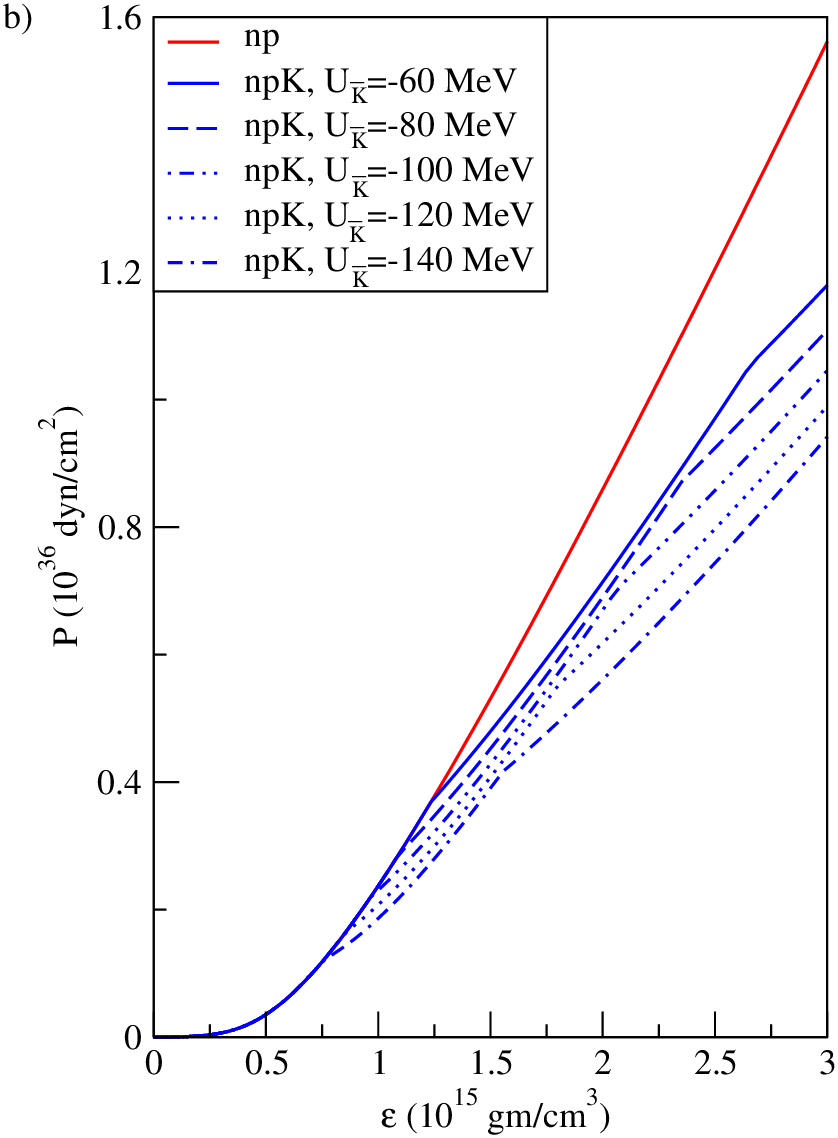}
\end{minipage}
\caption{Equation of state(EoS) for different compositions: a) np=red, 
\npl=blue, \nplsc=cyan, \nplc=green, \nplck=maroon colour in online version. 
b)np and npK for different $U_{\bar K}=-60$ to $-140$ MeV.}
\end{figure}
\begin{figure}
\begin{minipage}{.5\textwidth}
  \includegraphics[width=.8\textwidth, height=0.5\textheight]{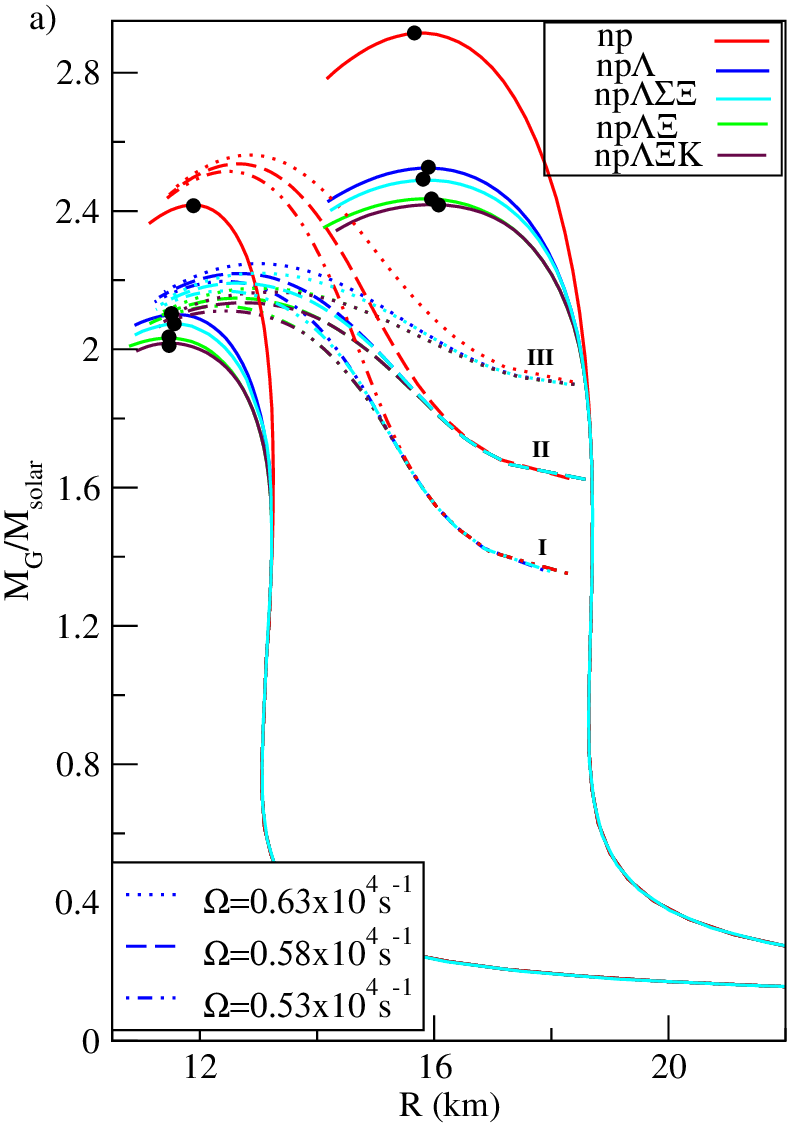}
\end{minipage}%
\begin{minipage}{.5\textwidth}
  \includegraphics[width=.8\textwidth,height=0.51\textheight]{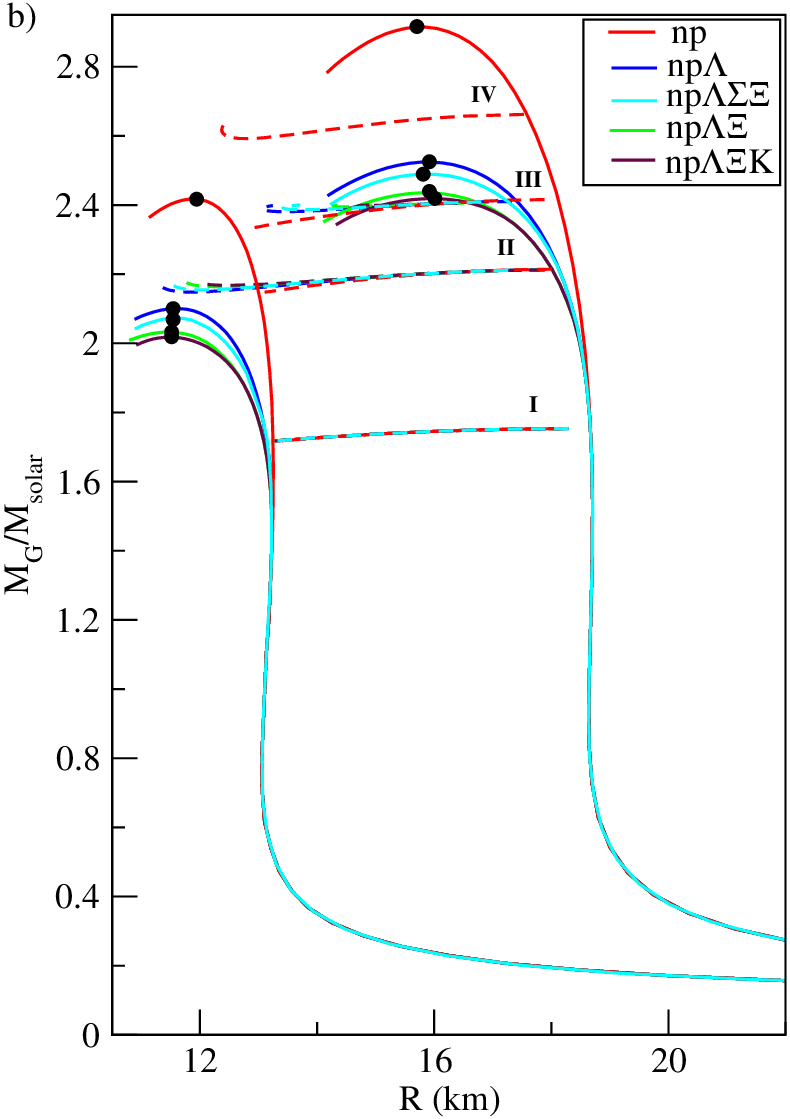}
\end{minipage}
\caption{Mass-radius profile for different compositions: np=red, \npl=blue, \nplsc=cyan,
\nplc=green, \nplck=maroon colour in online version. Static and Kepler 
mass-radius 
profiles are represented by solid lines on left and right extreme respectively. In between static and Kepler sequences are
a)different fixed angular frequency curves I, II and III for 
$\Omega$=5300,5800 and 6300 $s^{-1}$ respectively.
b)different rest mass curves I, II, III and IV for rest mass $M_R$= 1.92, 2.49, 2.75, and 3.23 $M_{solar}$ respectively.}
\end{figure}
\begin{figure}
\begin{minipage}{.5\textwidth}
  \includegraphics[width=.8\linewidth, height=0.5\textheight]{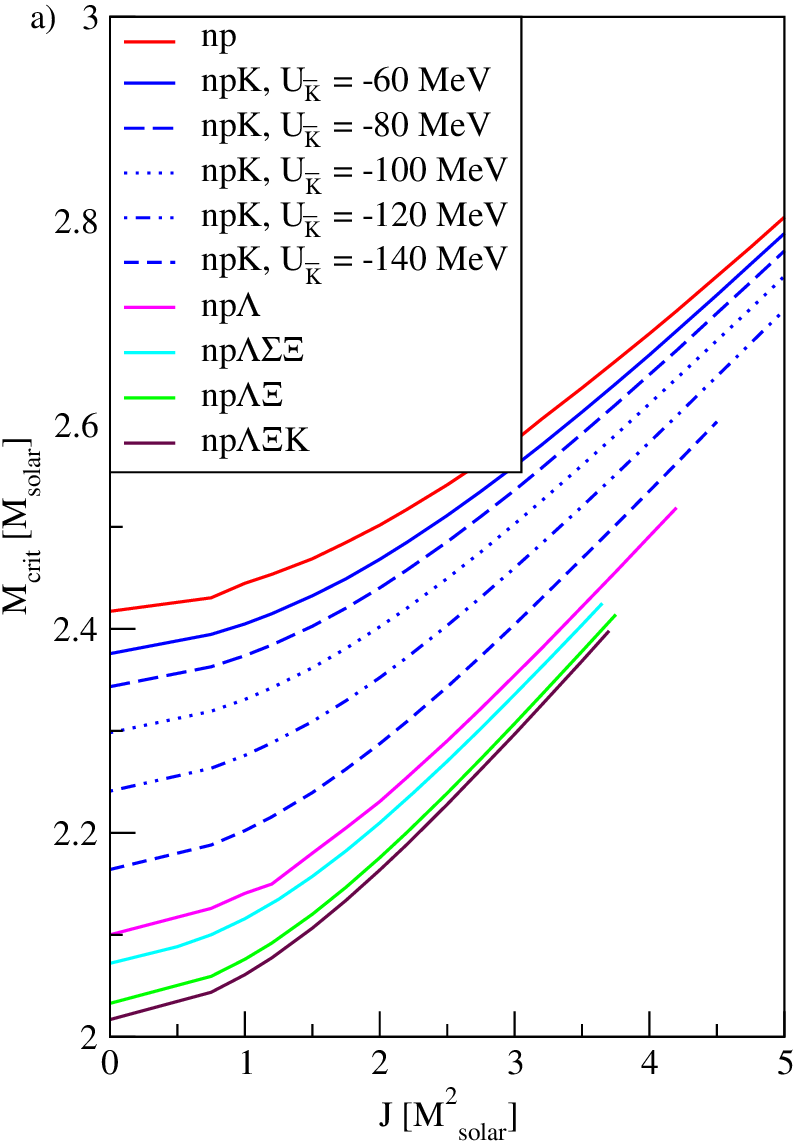}
\end{minipage}%
\begin{minipage}{.5\textwidth}
  \includegraphics[width=.8\linewidth,height=0.5\textheight]{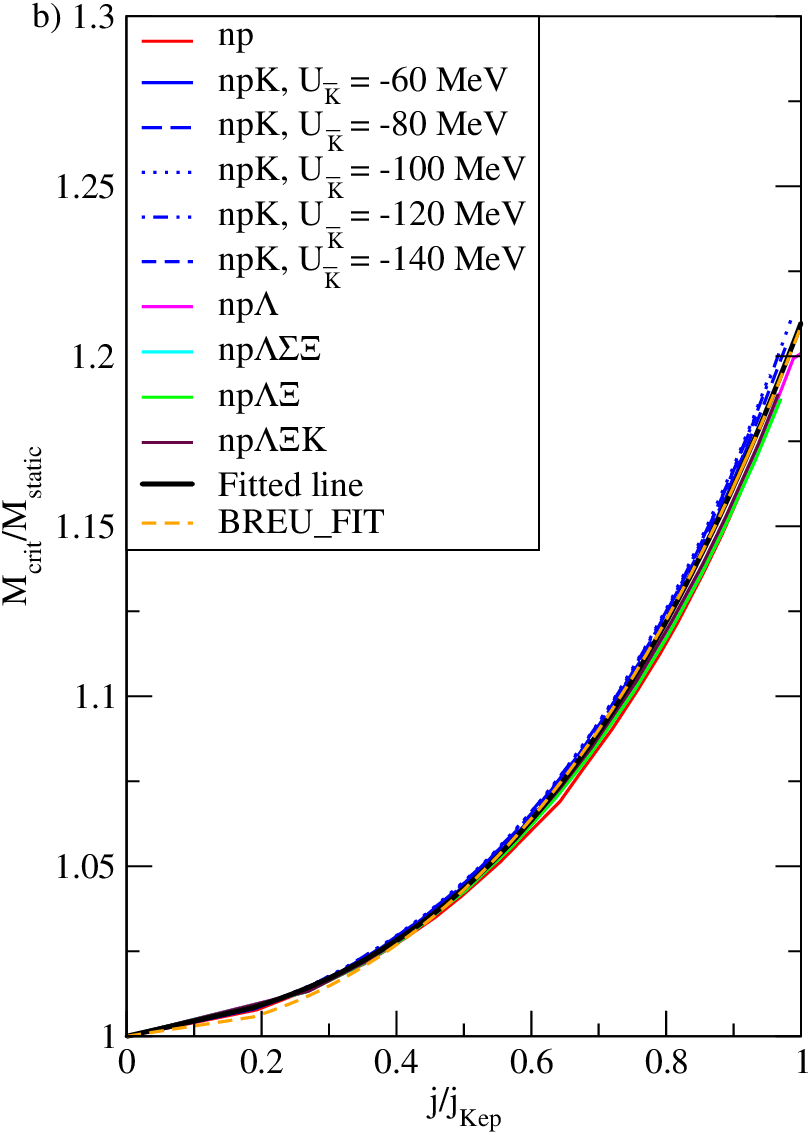}
\end{minipage}
  \caption{a) Critical mass i.e. the maximum mass of the constant angular
momentum stellar model versus the corresponding fixed angular momentum
  b)Normalised critical mass ($M_{crit}/M_{static}$) versus  normalised angular momentum.}
\end{figure}
\begin{figure}
\begin{minipage}{.5\textwidth}
  \includegraphics[width=.8\linewidth,height=0.5\textheight]{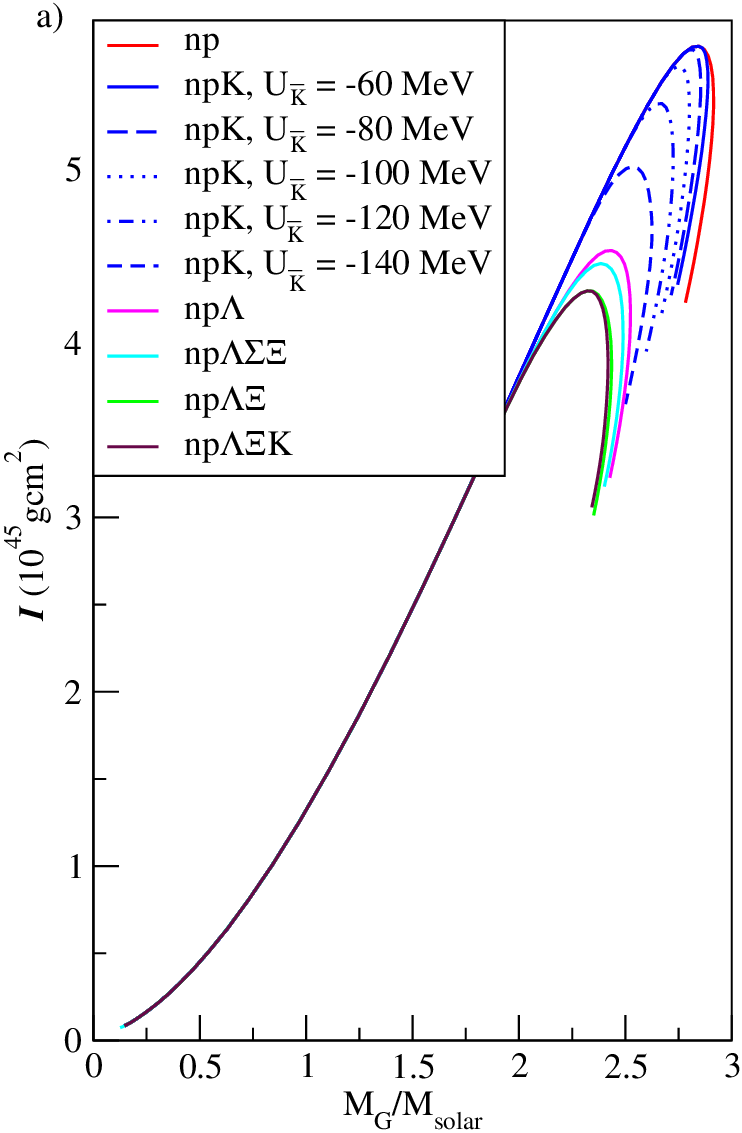}
\end{minipage}%
\begin{minipage}{.5\textwidth}
  \includegraphics[width=.8\linewidth,height=0.5\textheight]{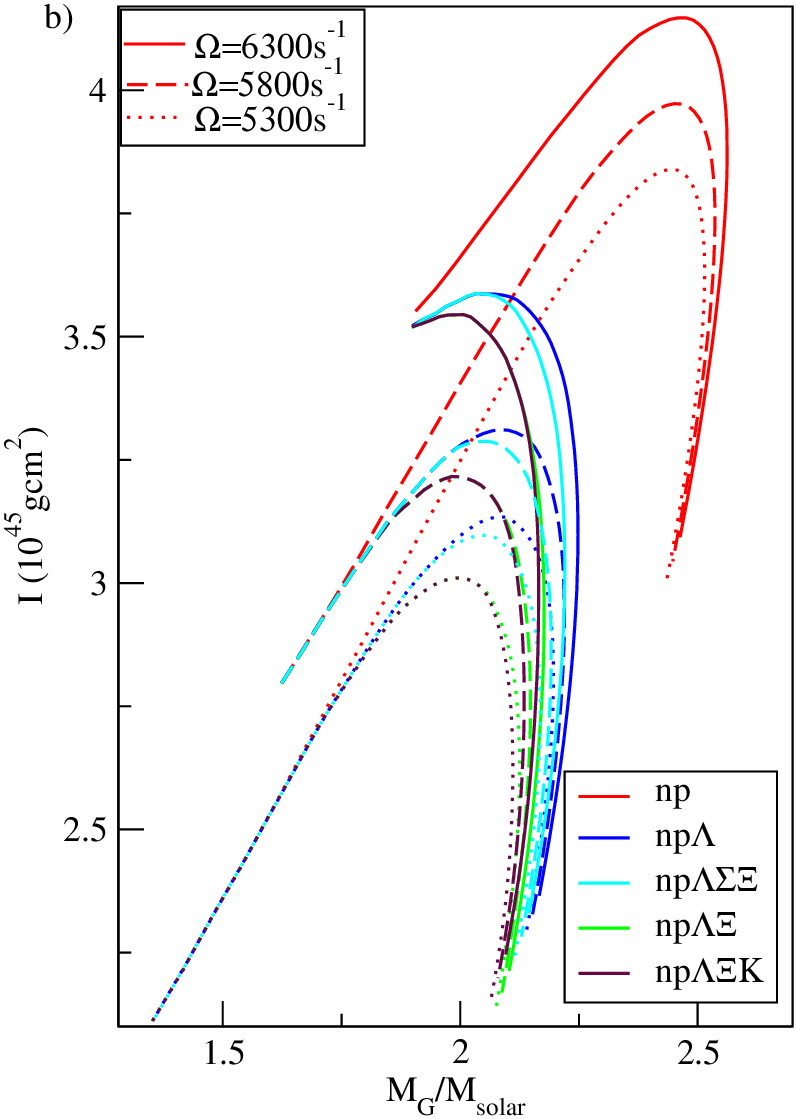}
\end{minipage}
\caption{Moment of inertia(\MI) versus gravitational mass($M_G$) for different 
compositions of the neutron stars rotating a)at Kepler frequency b)at different 
angular frequencies.}
\end{figure}
\begin{figure}
\begin{minipage}{.5\textwidth}
\centering
  \includegraphics[width=.8\textwidth,height=0.5\textheight]{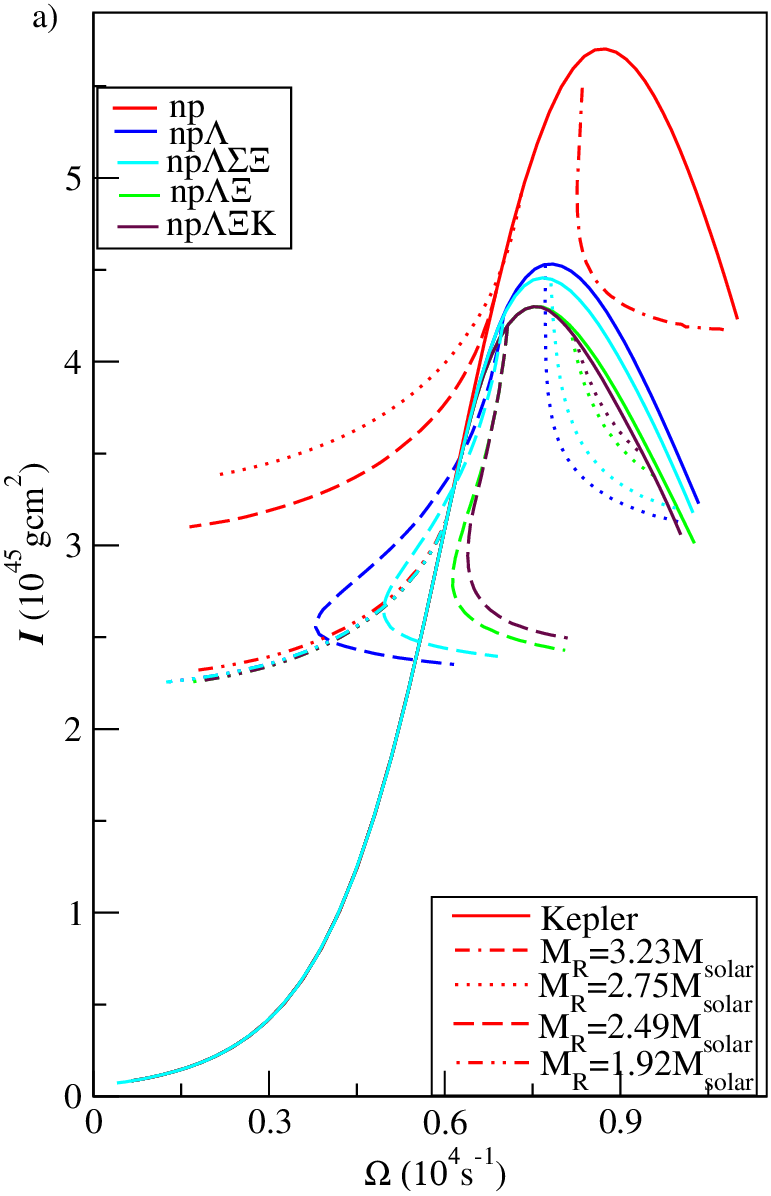}
\end{minipage}%
\begin{minipage}{.5\textwidth}
  \includegraphics[width=.8\textwidth,height=0.5\textheight]{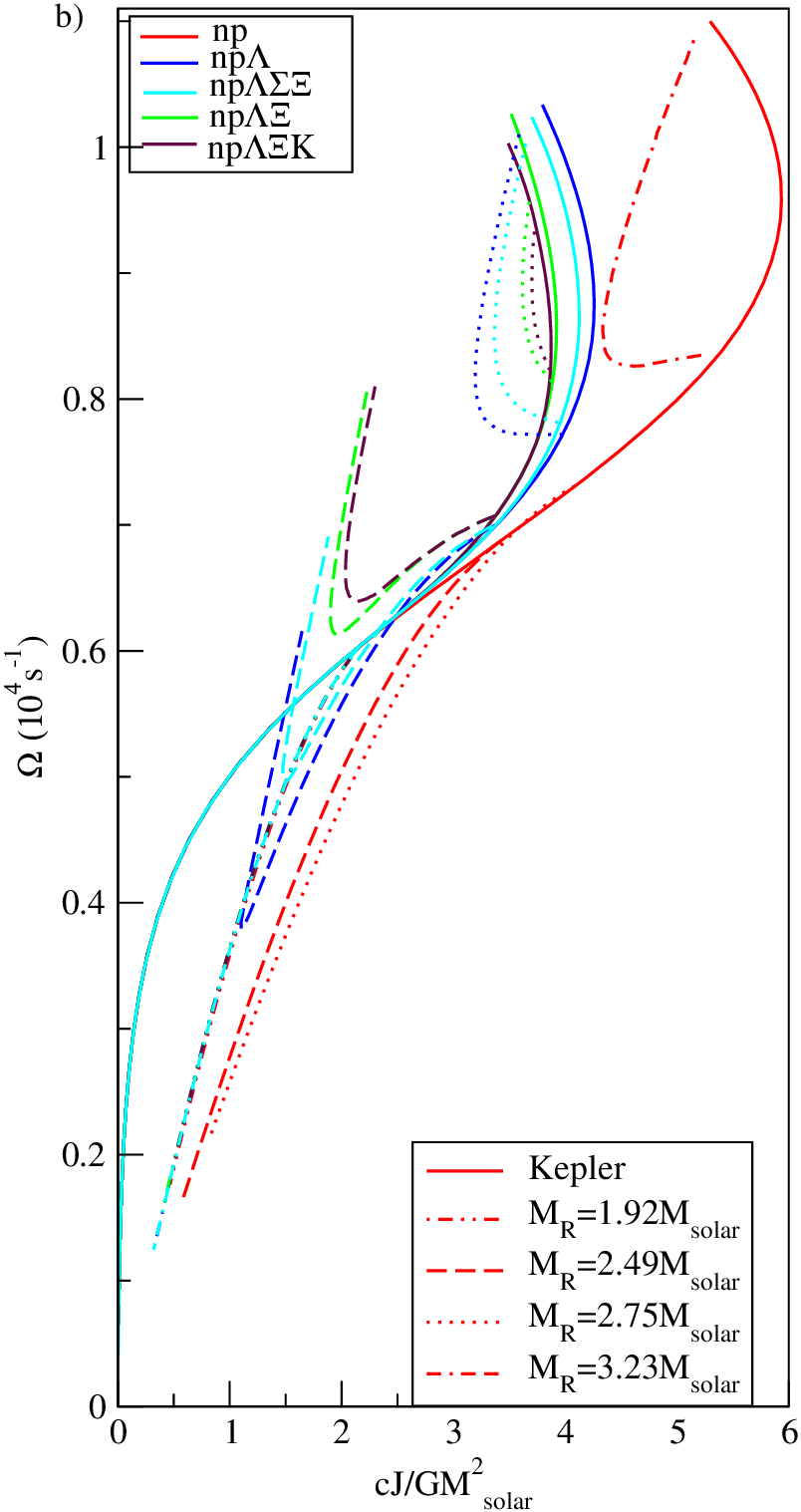}
\end{minipage}
  \caption{a)Moment of inertia(\MI) versus angular velocity($\Omega$). b)Angular velocity($\Omega$) versus angular momentum ($cJ/GM_{solar}^2$). The solid lines correspond to Kepler frequency. Others are for different fixed rest mass sequences.}
\end{figure}
\begin{figure}
\begin{minipage}{.5\textwidth}
  \includegraphics[width=.8\linewidth,height=0.5\textheight]{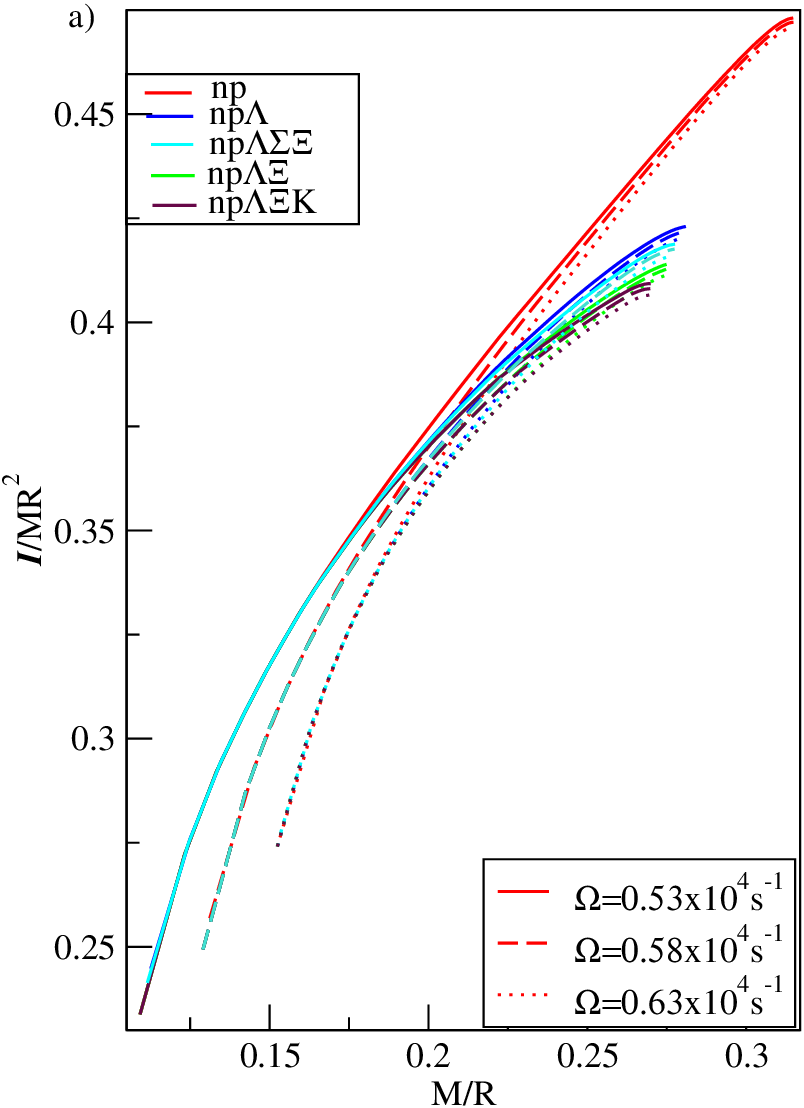}
\end{minipage}%
\begin{minipage}{.5\textwidth}
  \includegraphics[width=.8\linewidth,height=0.5\textheight]{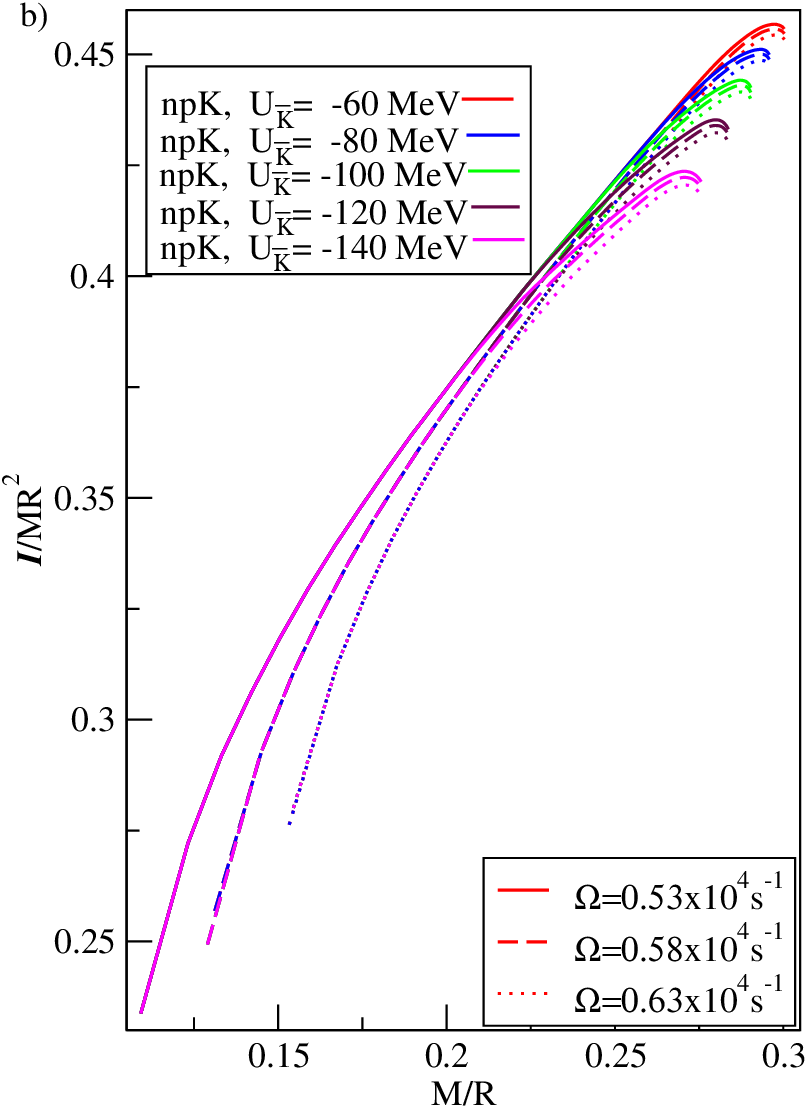}
\end{minipage}
\caption{Normalised moment of inertia(\MI$/MR^2$) with compactness for a)np, \npl, \nplsc, \nplc, \nplck, $U_{\bar K}=-140MeV$ b)\npk ~with different $U_{\bar K}$}
\end{figure}
\begin{figure}
\begin{minipage}{.5\textwidth}
  \includegraphics[width=.8\linewidth,height=0.65\textheight]{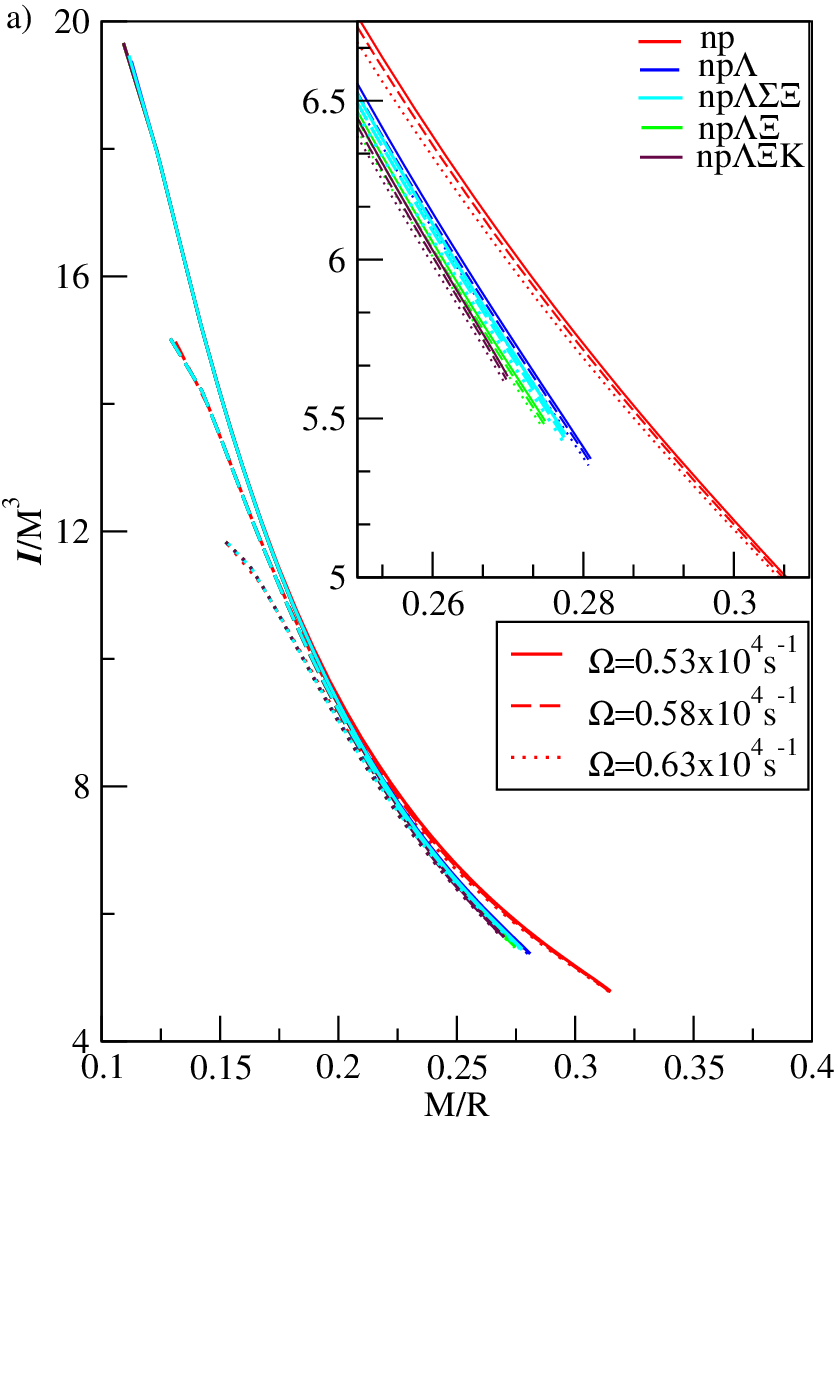}
\end{minipage}%
\begin{minipage}{.5\textwidth}
  \includegraphics[width=.8\linewidth,height=0.65\textheight]{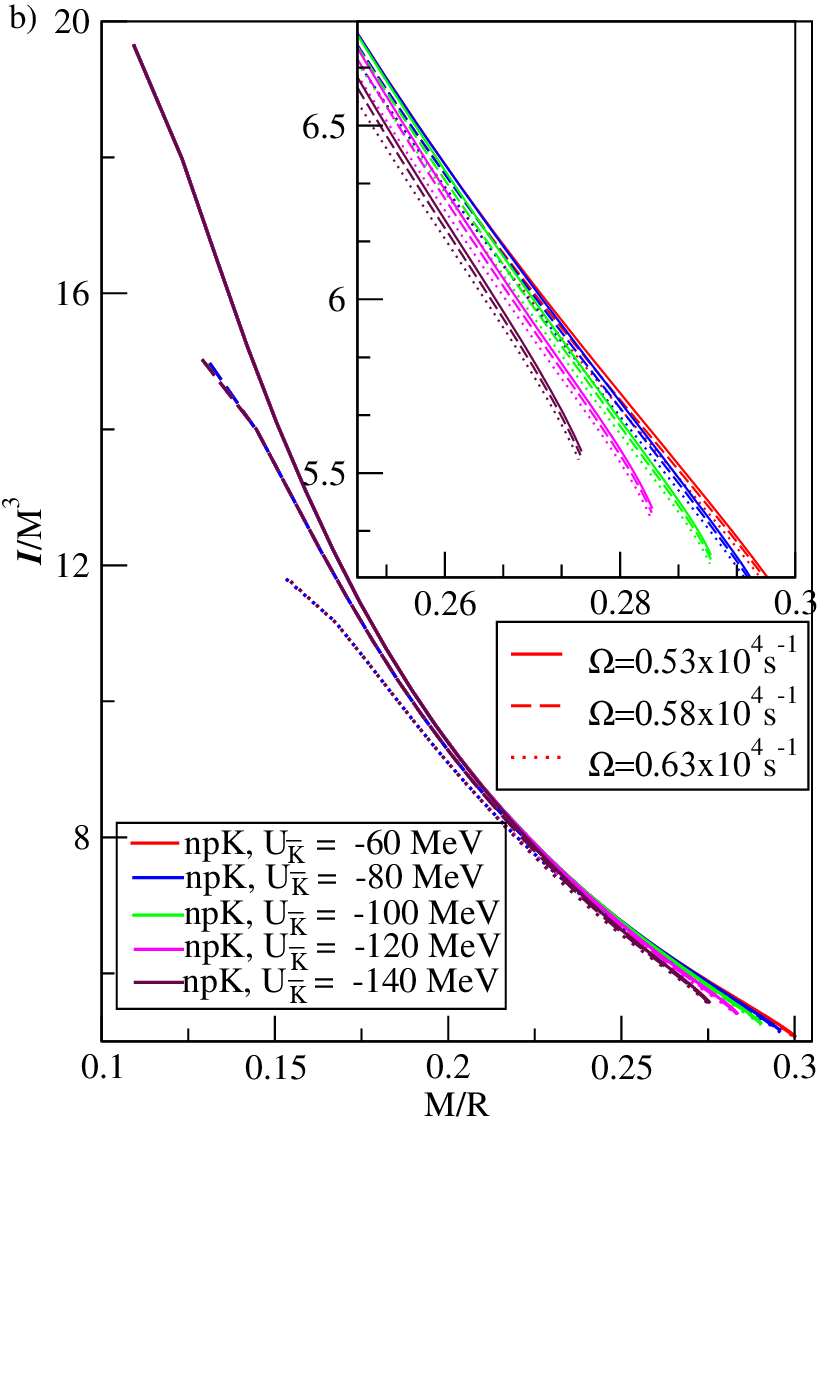}
\end{minipage}
\caption{Normalised moment of inertia(\MI$/M^3$) with compactness for a)np, \npl, \nplsc, \nplc, \nplck, $U_{\bar K}=-140MeV$ b) \npk ~with different $U_{\bar K}$}
\end{figure}
\begin{figure}
\begin{minipage}{.5\textwidth}
  \includegraphics[width=.8\linewidth,height=0.5\textheight]{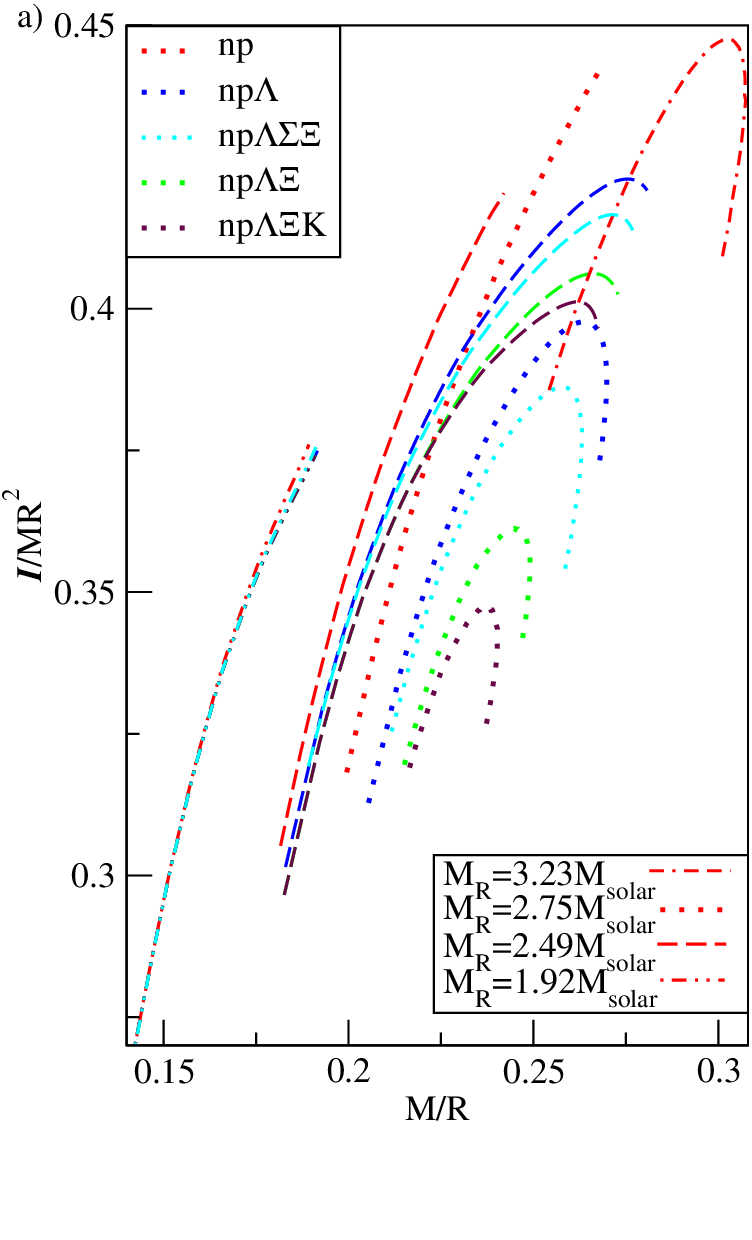}
\end{minipage}%
\begin{minipage}{.5\textwidth}
  \includegraphics[width=.8\linewidth,height=0.5\textheight]{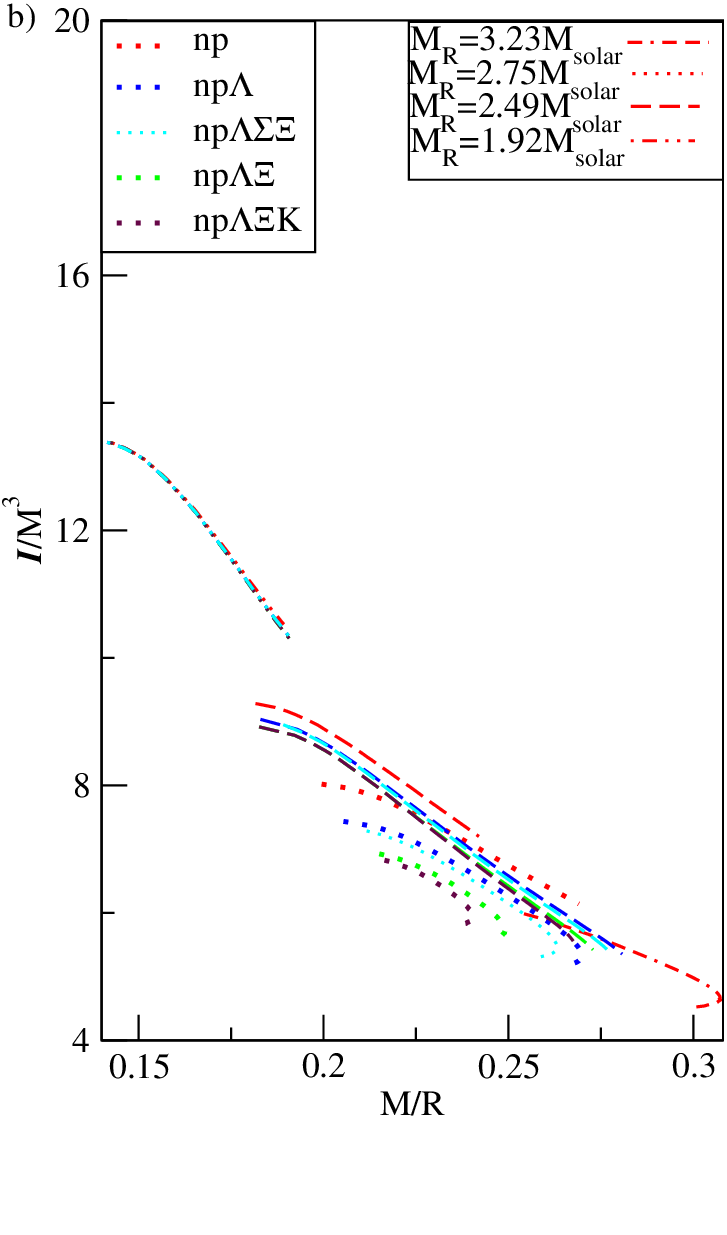}
\end{minipage}
\caption{Normalised moment of inertia as a function of compactness M/R a) \MI$/MR^2$ versus M/R b)\MI$/M^3$ versus M/R for the 
rest mass sequences $M_R=1.92$, $2.49$, $2.75$ and $3.23M_{solar}$.}

\end{figure}
\end{document}